\documentclass[aps,prd,preprint,showpacs,superscriptaddress,showkeys,floatfix,nofootinbib]{revtex4-1}
\usepackage{graphicx}
\usepackage{dcolumn}
\usepackage{bm}
\usepackage{color}
\usepackage{url}
\usepackage{amsmath,amssymb,graphicx}
\usepackage{amsmath}
\usepackage{amssymb}
\usepackage{mathrsfs}
\usepackage{accents}
\usepackage{epsfig}
\usepackage[lofdepth,lotdepth,caption=false]{subfig}
\usepackage[normalem]{ulem}
\usepackage{mathrsfs}
\usepackage{amsfonts}
\usepackage{feynmp}
\usepackage{multirow}
\usepackage{array}
\usepackage{slashed}
\usepackage{units}
\usepackage{bbm}
\usepackage{framed}
\usepackage[normalem]{ulem}
\usepackage{float}

\newlength\figureheight 
\newlength\figurewidth 

\begin{document}

\title{Hidden Interactions of Sterile Neutrinos As a Probe For New Physics}

\author{Zahra  Tabrizi}
\email{tabrizi.physics@ipm.ir}
\affiliation{School of Particles and Accelerators, Institute for Research in Fundamental Sciences (IPM), P.O.Box 19395-1795, Tehran, Iran}
\affiliation{Instituto de F\'isica Gleb Wataghin - UNICAMP, 13083-859, Campinas, SP, Brazil}
\author{O. L. G. Peres}
\email{orlando@ifi.unicamp.br}
\affiliation{Instituto de F\'isica Gleb Wataghin - UNICAMP, 13083-859, Campinas, SP, Brazil}
\affiliation{Abdus Salam International Centre for Theoretical Physics, ICTP, I-34010, Trieste, Italy}
\affiliation{School of Particles and Accelerators, Institute for Research in Fundamental Sciences (IPM), P.O.Box 19395-1795, Tehran, Iran}
\date{\today} 

\begin{abstract}
Recent results from neutrino experiments show evidence for light sterile neutrinos which do not have any Standard Model interactions. In this work we study the hidden interaction of sterile neutrinos with an "MeV scale" gauge boson (the $\nu_s$HI model) with mass $M_X$ and leptonic coupling $g^\prime_l$. By performing an analysis on the $\nu_s$HI model using the data of the MINOS neutrino experiment we find that the values above $G_X/G_F=92.4$ are excluded by more than $2\sigma$ C.L., where $G_F$ is the Fermi constant and $G_X$ is the field strength of the $\nu_s$HI model. Using this model we can also probe other new physics scenarios. We find that the region allowed by the $(g-2)_\mu$ discrepancy is entirely ruled out for $M_X\lesssim 100$ MeV. Finally, the secret interaction of sterile neutrinos has been to solve a conflict between the sterile neutrinos and cosmology. It is shown here that such an interaction is excluded by MINOS for $g^\prime_s> 1.6\times10^{-2}$. This exclusion, however, does depend on the value of $g_l^{\prime}$. 
\end{abstract}

\maketitle

\section{Introduction}
Most of the data collected from the neutrino oscillation experiments are in agreement with the 3 neutrino hypothesis~\cite{GonzalezGarcia:2002dz}. However, the observation of the reactor anomaly, which is a deficit of electron anti-neutrinos produced in the reactors~\cite{Mention:2011rk, *Huber:2011wv}, together with the results of the MiniBooNE experiment~\cite{AguilarArevalo:2010wv} which shows evidence for $\nu_{\mu}\to \nu_e$ conversion, cannot be explained by the usual 3 neutrino scenario~\cite{Abazajian:2012ys}. The most popular way to clarify these anomalies is to assume there exists 1 (or more) neutrino state(s) which does not have any weak interaction (therefore is sterile), but can mix with the active neutrinos in the Standard Model (SM) and change their oscillation behavior pattern.

Although most of the anomalies seen in the neutrino sector are in favor of the sterile models with mass $\sim1$ eV, there are conflicts between the sterile hypothesis and cosmology. Such light additional sterile states thermalize in the early universe through their mixing with the active neutrinos; therefore, we effectively have additional relativistic number of neutrinos which can be parametrized by $\Delta N_{\rm eff}$. In the standard model of cosmology we have $\Delta N_{\rm eff}=0$. Massive sterile neutrinos with mass $\sim1~$eV and large enough mixing angles to solve the reactor anomalies imply full thermalization at the early universe. This means that for any additional species of sterile neutrinos, we should have $\Delta N_{\rm eff}=1$. However, this is not consistent with the Big Bang Nucleosynthesis and the Planck results, which state $\Delta N_{\rm eff}<0.7$ with $90\%$ C.L.~\cite{Ade:2015xua}. It was recently proposed in \cite{Hannestad:2013ana,Dasgupta:2013zpn} that this problem could be solved if the sterile neutrino state interacts with a new gauge boson $X$ with mass $\sim$ a few MeV. This can easily produce a large field strength for the sterile neutrinos. In this way the sterile state experiences a large thermal potential which suppresses the mixing between the active and sterile states in the early universe. Therefore, the abundance of the sterile neutrinos remains small, and its impact on the Big Bang nucleosynthesis (BBN), Cosmic Microwave Background (CMB) and the Large Scale Structure Formation would be negligible, hence the sterile state can be consistent with the cosmological model. 

In this work we investigate the possibility of the sterile neutrino states interacting with a new gauge boson $X$, with mass $\sim$ MeV, which has couplings with the sterile neutrinos and the charged leptons in the SM. This new interaction of the sterile neutrinos was first mentioned in~\cite{Pospelov:2011ha}. The "\textbf{\textit{$\nu_s$~Hidden~Interaction}}" ($\nu_s$HI) model produces a neutral current (NC) matter potential for the sterile states proportional to $G_X$, where $G_X$ is the field strength of the new interaction. The NC matter potential in the $\nu_s$HI model changes the oscillation probability of neutrinos and anti-neutrinos drastically. Therefore, using the data of a neutrino oscillation experiment such as the MINOS experiment~\cite{Adamson:2013whj}, we can test the $\nu_s$HI model. 

An advantage of the $\nu_s$HI model is that through it we can use the data of neutrino oscillation experiments to test other new physics scenarios which imply having couplings with a light gauge boson, such as the explanation of the $(g-2)_\mu$ discrepancy with a light gauge boson \cite{Pospelov:2008zw} and the secret interaction of sterile neutrinos proposed in \cite{Hannestad:2013ana,Dasgupta:2013zpn} which solves the tension between the sterile hypothesis and cosmology. 

\section{The Formalism}
We enlarge the SM with one extra species of the sterile neutrinos which do not couple with the SM gauge bosons, but have interactions with a new $U_X(1)$ gauge symmetry (the {$\nu_s$HI model}). {The new gauge boson couples to the sterile neutrinos and charged leptons with coupling constants $g^{\prime}_s$ and $g^{\prime}_l$, respectively~\cite{footnote1}}, where for simplicity, we have assumed equal coupling constants for the charged leptons. The strength of this new interaction is given by 
\begin{equation}\label{eq4.1}
\frac{G_X}{\sqrt{2}}=\frac{g^{\prime}_sg^{\prime}_l}{4M_X^2},
\end{equation}
where $M_X$ is the mass of the new gauge boson.

The active neutrinos of the SM have charged and neutral current interactions with the  $W^{\pm}$ and $Z$ bosons. Their matter potential is therefore given by $V_{\alpha}(r)=\delta_{\alpha e}V_{CC}(r)+V_{NC}(r)=\sqrt{2}G_FN_e(r)(\delta_{\alpha e}-1/2)$, where $\alpha=e,\mu,\tau$ and $V_{CC(NC)}$ is the charged (neutral) current potential of the active neutrinos. The factor $G_F$ is the Fermi constant, while $N_e(r)$ is electron number density of the earth given by the PREM model~\cite{Dziewonski1981297}. We have assumed that the electron and neutron number densities are equal for our practical purposes.

The sterile neutrinos which couple to the $X$ boson will also have neutral current matter potential which is proportional to the strength field of the new interaction:
\begin{eqnarray}\label{eq4.2}
V_s(r)=-\frac{\sqrt{2}}{2}G_XN_e(r)\equiv\alpha V_{NC}(r),
\end{eqnarray}
where the dimensionless parameter $\alpha$ is defined as
\begin{equation}\label{eq4.3}
\alpha=\frac{G_X}{G_F}.
\end{equation}
For $\alpha\to0$ we recover the minimal $3+1$ sterile neutrino model. In the minimal $"3+1"$ model~\cite{Peres:2000ic} the flavor and mass eigenstates of neutrinos are related through the unitary $(3+1)\times(3+1)$ PMNS matrix $U$: $\nu_{\alpha}=\sum_{i=1}^{3+1}U^*_{\alpha i}\nu_i$. The oscillation probability of neutrinos is described using the active-active and active-sterile mixing angles, as well as the mass squared differences $\Delta m^2_{21}$, $\Delta m^2_{31}$ and $\Delta m^2_{41}$, where $\Delta m^2_{ij}\equiv m_i^2-m_j^2$.

The evolution of neutrinos in the $\nu_s$HI model can be found by solving the following Schr\"{o}dinger-like equation
\begin{eqnarray}\label{eq4.4}
i\frac{d}{dr}
\begin{pmatrix}
\nu_e\\\nu_\mu\\\nu_\tau\\\nu_s
\end{pmatrix}
=\Big[\frac{1}{2E_\nu}UM^2{U}^\dagger+V^{\nu_s{\rm{SI}}}(r)\Big]
\begin{pmatrix}
\nu_e\\\nu_\mu\\\nu_\tau\\\nu_s
\end{pmatrix},
\end{eqnarray}
where $U$ is the $4\times4$ PMNS matrix~\cite{Abazajian:2012ys}, which is parametrized by the active-active mixing angles $(\theta_{12},\theta_{13},\theta_{23})$ as well as 3 active-sterile mixing angles $(\theta_{14},\theta_{24},\theta_{34})$. 
The matrix 
$$M^2=\rm{diag}\Big(0,\Delta m^2_{21},\Delta m^2_{31},\Delta m^2_{41}\Big)$$
 is the matrix of the mass squared differences. Using Eq.~(\ref{eq4.2}), the matter potential matrix in the $\nu_s$HI model will be (after subtracting the constant $V_{NC}(r)\times\mathbb{I}$)
\begin{eqnarray}\label{eq4.5}
V^{\nu_s\rm{HI}}(r)&=&{\rm diag}\Big(V_{\rm{CC}}(r),0,0,V_s(r)-V_{\rm{NC}}(r)\Big)\nonumber\\
&=&\sqrt{2}G_FN_e(r)\rm{diag}\Big(1,0,0,\frac{(1-\alpha)}{2}\Big). 
\end{eqnarray}
The same evolution equation applies to anti-neutrinos with the replacement $V^{\nu_s\rm{HI}}(r)\to-V^{\rm{\nu_sHI}}(r)$. We consider the $\nu_s$HI model with $\alpha>0$.  In an effective 2-neutrino scheme the so called  MSW resonance~\cite{msw} happens when $\frac{\Delta m^2}{2E_\nu}\cos\theta=V$. Since in the $\nu_s$HI model the sterile states have nonzero matter potential, the potential would be positive in a $\nu_\mu-\nu_s$ system (for $\alpha>1$), which means that at energies where the resonance condition is carried out, $\nu_\mu$ converts to $\nu_s$.

An interesting place to test the $\nu_s$HI model is the MINOS long-baseline neutrino experiment~\cite{Adamson:2013whj}. The MINOS experiment which has a baseline of $735$~km detects both muon and anti-muon neutrinos, and it is one of few experiments that is both sensitive to neutrino and anti-neutrino oscillation probabilities. For the baseline and energy range of the MINOS experiment, the oscillation probabilities of the neutrinos and anti-neutrinos are very similar in the usual 3 neutrino scenario. However, this does not hold in the $\nu_s$HI model anymore. 

\begin{figure}[!hHtb]
\includegraphics[scale=1]{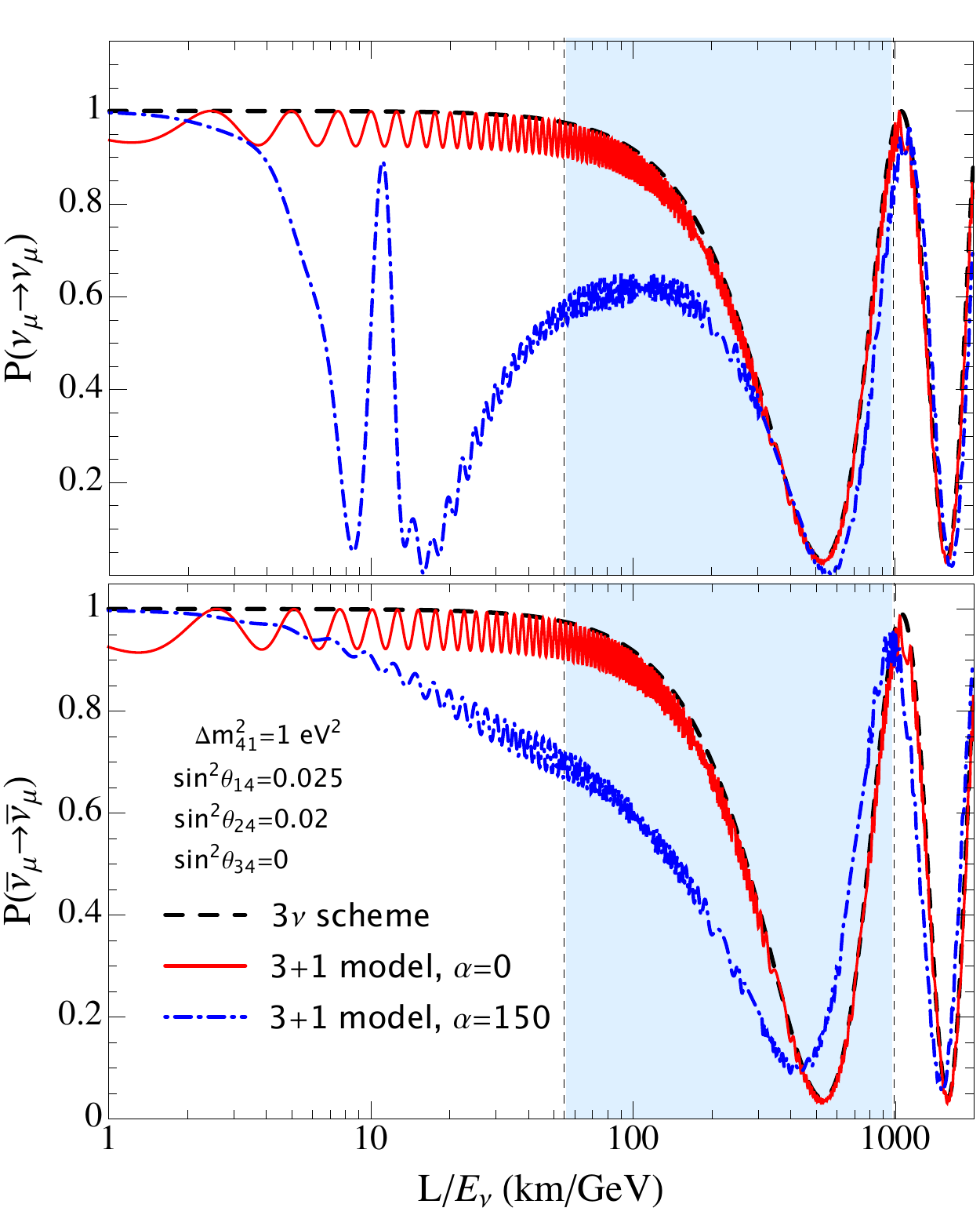}
\caption{\label{probability}
The muon neutrino and anti-neutrino survival probabilities as a function of distance over neutrino energy are shown in the top and bottom, respectively. The black-dashed and the red-solid curves correspond to the 3 and $3+1$ neutrino models, respectively. The blue dot-dashed curves represent the probabilities calculated in the $\nu_s$HI model for $\alpha=150$. The standard 3 neutrino parameters are fixed by the NUFIT best fit values~\cite{GonzalezGarcia:2012sz} and the active-sterile mixing parameters are shown in the plot. The blue shaded area is the range of $L/E_\nu$ for the MINOS experiment~\cite{Adamson:2013whj}.
}
\end{figure}

To see how the $\nu_s$HI model affects the oscillation probability of neutrinos, we compute the full numerical survival probabilities for muon (anti-)neutrino in the case of the standard 3 neutrino scenario and in the $3+1$ and $\nu_s$HI models. We show in Fig.~(\ref{probability}) the survival probability of $\nu_\mu$ (top) and $\bar{\nu}_\mu$ (bottom) for the standard 3 neutrino case (black dashed curve), the $3+1$ model with $\alpha=0$ (red solid curve) and the $\nu_s$HI model with $\alpha=150$ (the blue dot-dashed curve). To calculate the probabilities, we have fixed the 3 neutrino oscillation parameters by the best fit values of NUFIT~\cite{GonzalezGarcia:2012sz}: 
$\Delta m^2_{21}=7.5\times 10^{-5} \hspace{2pt}\mathrm{eV}^2$,
$\Delta m^2_{31}=2.4\times 10^{-3} \hspace{2pt}\mathrm{eV}^2$, $\sin^2\theta_{12}=0.3$,
$\sin^2\theta_{23}=0.6$  and $\sin^2\theta_{13}=0.023$. The values of the active-sterile mixing parameters in the 3+1 and $\nu_s$HI models are listed in Figure~(\ref{probability})~\cite{Abazajian:2012ys}. 
Comparing the 3 neutrino case (the black dashed curve) with the $3+1$ model (the red solid curve), we see that the effect of the sterile neutrino with mass squared difference $\Delta m_{41}^2=1$~eV$^2$ is marginal adding only a very fast oscillation on the top of the oscillation induced by the atmospheric mass squared difference $\Delta m_{31}^2$. However, in the  $\nu_s$HI model we have dramatic effects both for neutrino and anti-neutrino survival probabilities. When the resonance condition is fulfilled, we expect stronger changes for the $\nu_\mu$ survival probability, while for anti-neutrinos the changes are milder. This can be seen in the blue dot-dashed curve at the top and bottom of Fig.~(\ref{probability}).

\section{The~Analysis}
We analyze the collected $\nu_\mu$ and $\bar{\nu}_\mu$ beam data in the MINOS experiment to constrain the $\alpha$ parameter in the $\nu_s$HI model. We calculate the expected number of events in each bin of energy by
\begin{eqnarray}\label{eq4.6}
N^{{\rm osc}}_i
=N^{\rm no-osc}_i\times \left<P_{{\rm sur}}(s_{23}^2,s_{24}^2,\Delta m^2_{31},\Delta m^2_{41};\alpha)\right>_i,
\end{eqnarray}
where $s_{ij}^2\equiv\sin^2\theta_{ij}$ and $N^{{\rm no-osc}}_i$ is the expected number of events for no-oscillation case in the $i$th bin of energy after subtracting background~\cite{Adamson:2013whj};  while $\left<P_{{\rm sur}}\right>_i$ is the averaged $\nu_{\mu}\to\nu_{\mu} ~(\bar{\nu}_{\mu}\to\bar{\nu}_{\mu})$ survival probability in the $i$th energy bin, calculated using Eq.~(\ref{eq4.4}) for the fixed values  $\sin^2\theta_{14}=0.025$ and $\sin^2\theta_{34}=0$ and letting the other parameters to vary.

To analyze the full MINOS data we define the following $\chi^2$ function
\begin{eqnarray}\label{eq4.7}
\chi^2=\sum_i\frac{\Big[(1+a)N^{{\rm osc}}_i+(1+b)N_i^b-N_i^{{\rm obs}}\Big]^2}{(\sigma_i^{\rm obs})^2}+\frac{a^2}{\sigma_a^2}+\frac{b^2}{\sigma_b^2},\nonumber\\
\end{eqnarray}
where $i$ runs over the bins of energy (23 for $\nu_\mu$ events and 12 for $\bar{\nu}_\mu$ events), $N^{{\rm osc}}_i$ is the expected number of events defined in Eq.~(\ref{eq4.6}), $N_i^b$ and $N_i^{\rm{obs}}$ are the background and observed events, respectively. The $\sigma_i^{\rm obs}=\sqrt{N_i^{\rm{obs}}}$ represents the statistical error of the observed events. The parameters $a$ and $b$ take into account the systematic uncertainties of the normalization of the neutrino flux and the background events respectively, with $\sigma_a=0.016$ and $\sigma_b=0.2$~\cite{footnote2}. 

\begin{figure}[!hHtb]
\includegraphics[scale=0.7]{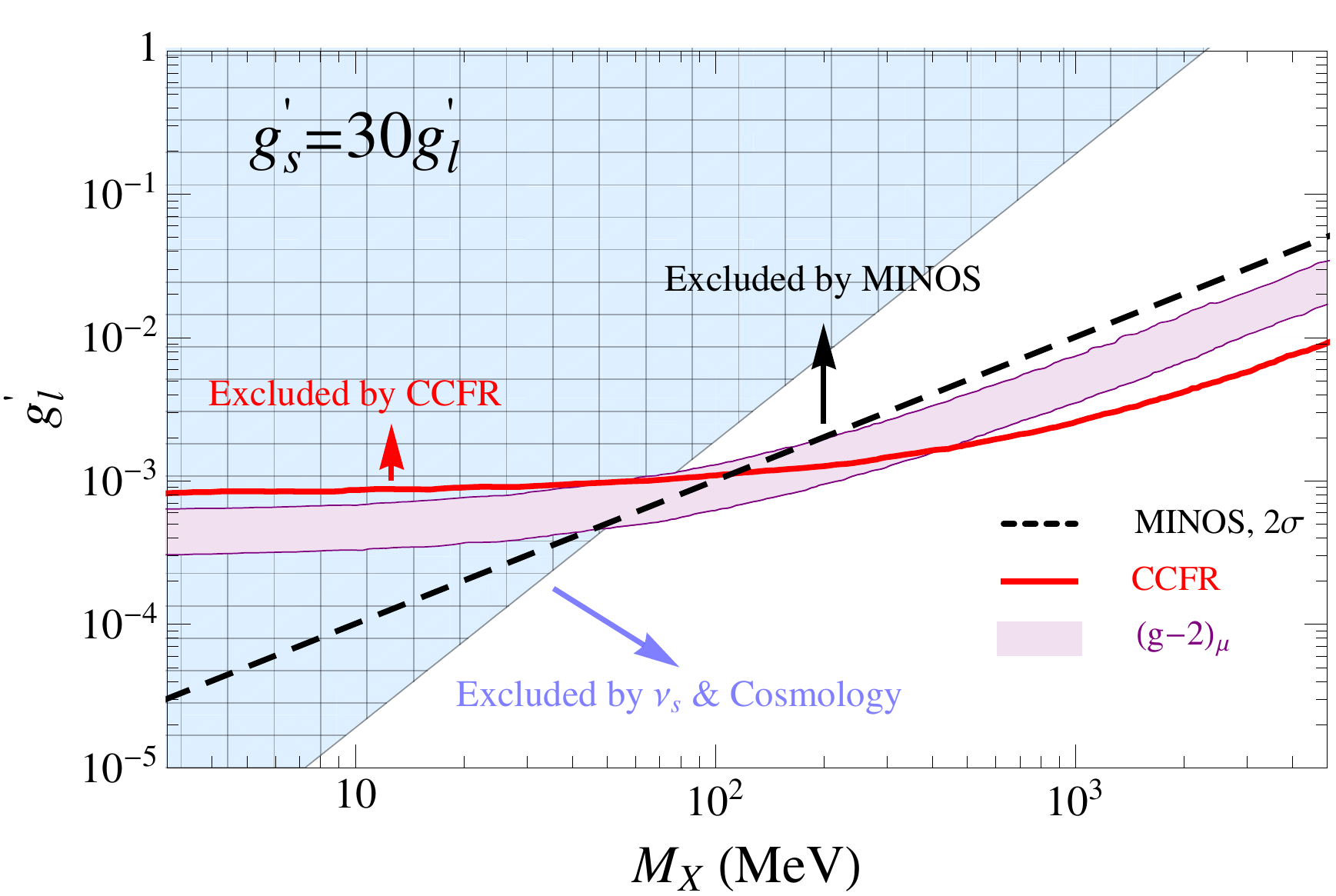}
\caption{\label{gx-mx}
We have shown the region of interest for the $\nu_s$HI model with a light gauge boson with mass $M_X$ and couplings $g^\prime_l$ and $g^\prime_s=\gamma g^\prime_l$. The result of the analysis of the $\nu_s$HI model with the MINOS data is shown by the black dashed curve with $2\sigma$ C.L. (for $\gamma=30$). The purple shaded region is the region favored by the $(g-2)_\mu$ discrepancy, while the red curve is the CCFR~\cite{Mishra:1991bv} measurement of the neutrino trident cross-section~\cite{Altmannshofer:2014pba}. The blue shaded region is where the tension between the sterile neutrino and cosmology is relieved for $f(g^\prime_s,M_X)=100$ and $\gamma=30$ (See Eq.~(\ref{eq4.10}) and the discussion after that). 
}
\end{figure}

After combining the $\chi^2$ function for the $\nu_\mu$ and $\bar{\nu}_\mu$ events and marginalizing over all parameters, we find the following best fit values: 
$\Delta m^2_{31}=2.43\times10^{-3}~\rm{eV}^2$, $\Delta m^2_{41}=4.35~\rm{eV}^2$, $s_{23}^2=0.67$, $s^2_{24}=0.03$,  and $\alpha=19.95$, while the ratio of the $\chi^2$ value over the number of degrees of freedom is $\chi^2/{\rm{d.o.f}}=39.7/30$.  When we increase $\alpha$ from its best fit value we have disagreement between the $\nu_s$HI model and the MINOS data.  From this we can find an upper bound for $\alpha$ at $2\sigma$ C.L.:
\begin{eqnarray}\label{eq4.8}
\alpha<92.4.
\end{eqnarray}
Using Eq.~(\ref{eq4.1}) and Eq.~(\ref{eq4.3}), we can write down the coupling $g^{\prime}_l$ as a function of the gauge boson mass $M_X$ and fixed value of $\alpha$:
\begin{eqnarray}\label{eq4.9}
g^{\prime}_l=\sqrt{\frac{2\sqrt{2}\alpha G_F}{\gamma}}M_X=5.5\times10^{-5}\sqrt{\frac{\alpha}{92.4\gamma}}
(\frac{M_X}{{\rm MeV}} ),
\end{eqnarray}
where we have assumed the two new coupling constants in our model are related as $g^{\prime}_s=\gamma g^{\prime}_l$, in which $\gamma\ge1$. Therefore, we can use the expression above to find an exclusion region in the $(M_X-g^\prime_l)$ plane. Implementing the relation above for the MINOS experiment, we arrive to the the black dashed curve shown in Fig.~(\ref{gx-mx}).

A light gauge boson with mass $\sim$ MeV can be used as a unique explanation for the $3.6~\sigma$ discrepancy between the experimental measurement and the SM prediction of the muon anomalous magnetic moment, $(g-2)_\mu$ \cite{Pospelov:2008zw}. The purple shaded region in Fig.~(\ref{gx-mx}) shows the favored $2\sigma$ region from $(g-2)_\mu$ discrepancy. It is shown in Ref.~\cite{Davoudiasl:2014kua} that nearly the entire $(g-2)_\mu$ band is excluded by various experiments if one assumes that the light gauge boson decays to charged leptons with branching ratio (Br) $\sim1$. However, in the $\nu_s$HI model, the primary decay mode of the light gauge boson is into invisibles (such as the light sterile neutrinos) with Br~$\sim1$. Therefore, all the $(g-2)_\mu$ band in Fig.~(\ref{gx-mx}) will be valid in the $\nu_s$HI model. As Fig.~(\ref{gx-mx}) shows, by comparing our results on the MINOS analysis of the $\nu_s$HI model with the $(g-2)_\mu$ band we can exclude all the masses below $M_X\sim 100 \sqrt{\gamma/30}$~MeV with $2\sigma$ C.L.. 

Another piece of information comes from neutrino trident production: the process in which the $\mu^+\mu^-$ pair is produced from the scattering of $\nu_\mu$ off the Coulomb field of a nucleus.  The red solid curve in Fig.~(\ref{gx-mx}) represents the results of the constrains from CCFR experiment on measurement of  the neutrino trident cross-section~\cite{Mishra:1991bv}. As it can be seen from Fig.~(\ref{gx-mx}), by combining the result of the CCFR experiment with our result from the MINOS analysis, there is only a tiny region in the $(g-2)_\mu$ band which is allowed by all experiments.

The sterile neutrino states with 1 eV mass have dramatic effects in cosmology due to their thermalization in the early universe and are disfavored by the Planck data. This tension could be removed if the sterile states have interactions with a light gauge boson in the so called secret interaction model~\cite{Dasgupta:2013zpn,Kopp:2014fha}. This will produce a temperature dependent matter potential for the sterile states which is $V_{\rm{eff}}=-\frac{7\pi^2}{45}\frac{{g^\prime_s}^2}{M_X^4}E_{\nu}T_s^4$~\cite{Dasgupta:2013zpn}, where $T_s$ is the temperature of the sterile sector and $E_{\nu}\ll M_X$. Therefore, the oscillation of the active to sterile neutrinos would be suppressed if $|V_{\rm{eff}}|\gg|  \frac{\Delta m^2_{41}}{2E_{\nu}}|$~\cite{Dasgupta:2013zpn}. We define the following function: 
\begin{eqnarray}\label{eq4.10}
\dfrac{V_{\rm{eff}}}{\Delta m^2_{41}/2E_\nu} \equiv f(g^\prime_s,M_X)=\frac{14\pi^2 {g^\prime_s}^2 E_\nu^2}{45\Delta m^2_{41}}(\frac{T_s}{M_X})^4.
\end{eqnarray}
Hence, the cosmology condition in the secret interaction model would be satisfied if $f(g^\prime_s,M_X)\gg1$. Similar to Eq.~(\ref{eq4.9}) we can find the values of the coupling constant which satisfy the cosmology condition:
\begin{eqnarray}\label{eq4.11}
g^\prime_s\gg \sqrt{\frac{45}{14\pi^2}}\frac{\sqrt{\Delta m^2_{41}}}{ E_\nu}\Big(\frac{M_X}{T_s}\Big)^2.
\end{eqnarray}
Assuming that the cosmology condition is satisfied for $f(g^\prime_s,M_X)=100$, then using Eq.~(\ref{eq4.9}) and the relation between the 2 coupling constants $g^\prime_s=\gamma g^\prime_l$, the values of $g^\prime_s$ above
\begin{eqnarray}\label{eq4.12}
g^\prime_s=1.6\times10^{-2}\big(\frac{T_s}{\rm{MeV}}\big)^2\big(\frac{E_\nu/\rm{MeV}}{\sqrt{\Delta m^2_{41}/\rm{eV}^2}}\big)\frac{\alpha}{92.4}\frac{\gamma}{30}
\end{eqnarray}
 is excluded by the MINOS analysis.
Therefore, using the MINOS data we find that at the time of BBN ($E_{\nu}\simeq T_s\simeq 1$~MeV) and for $\Delta m^2_{41}=1~$eV$^2$, the values of the coupling constant above $g^\prime_s= 1.6\times10^{-2}$ is excluded with more than $2\sigma$ C.L. (for $\gamma=30$). The blue shaded region in Fig.~(\ref{gx-mx}) shows the cosmology condition for the values mentioned above.

\section{Conclusions}
We have investigated the possibility that the light sterile neutrinos as suggested by the reactor anomaly have hidden interaction with an "MeV scale" gauge boson. In the Hidden Interaction ($\nu_s$HI) model, the sterile neutrinos have neutral current matter potential. Therefore, we can use the data of the neutrino experiments to constrain this model and probe other new physics scenarios. The field strength of this model is described by $G_X$. In this work we studied the $\nu_s$HI model using the MINOS experiment and showed that the values above $G_X/G_F=92.4$ are excluded.

One consequence of the $\nu_s$HI model is constraining other new physics scenarios such as explaining the $(g-2)_\mu$ discrepancy with a light gauge boson. We showed that using the $\nu_s$HI model, the $(g-2)_\mu$ region is entirely ruled out for $M_X\lesssim 100 \sqrt{\gamma/30}$ MeV by the MINOS data.  Also, the secret interaction of sterile neutrinos which is introduced in the literature to solve the tension between the sterile neutrinos and cosmology is excluded by MINOS for $g^\prime_s> 1.6\times10^{-2}\frac{\gamma}{30}$ for any value of $M_X$, where $g^\prime_s$ is the coupling between the sterile states and the light gauge boson. We can use the data of the future neutrino oscillation experiments such as DUNE~\cite{Berryman:2015nua} to further test the $\nu_s$HI model and get a definite answer on the presence of the light gauge boson.

\section{Acknowledgments}
Z. T. thanks the the useful discussions with Joachim Kopp and Pedro Machado. O. L. G. P. thanks the hospitality of IPM and the support of FAPESP funding grant 2012/16389-1.

\bibliography{nusHI.bib}

\begin{thebibliography}{22}%
\makeatletter
\providecommand \@ifxundefined [1]{%
 \@ifx{#1\undefined}
}%
\providecommand \@ifnum [1]{%
 \ifnum #1\expandafter \@firstoftwo
 \else \expandafter \@secondoftwo
 \fi
}%
\providecommand \@ifx [1]{%
 \ifx #1\expandafter \@firstoftwo
 \else \expandafter \@secondoftwo
 \fi
}%
\providecommand \natexlab [1]{#1}%
\providecommand \enquote  [1]{``#1''}%
\providecommand \bibnamefont  [1]{#1}%
\providecommand \bibfnamefont [1]{#1}%
\providecommand \citenamefont [1]{#1}%
\providecommand \href@noop [0]{\@secondoftwo}%
\providecommand \href [0]{\begingroup \@sanitize@url \@href}%
\providecommand \@href[1]{\@@startlink{#1}\@@href}%
\providecommand \@@href[1]{\endgroup#1\@@endlink}%
\providecommand \@sanitize@url [0]{\catcode `\\12\catcode `\$12\catcode
  `\&12\catcode `\#12\catcode `\^12\catcode `\_12\catcode `\%12\relax}%
\providecommand \@@startlink[1]{}%
\providecommand \@@endlink[0]{}%
\providecommand \url  [0]{\begingroup\@sanitize@url \@url }%
\providecommand \@url [1]{\endgroup\@href {#1}{\urlprefix }}%
\providecommand \urlprefix  [0]{URL }%
\providecommand \Eprint [0]{\href }%
\providecommand \doibase [0]{http://dx.doi.org/}%
\providecommand \selectlanguage [0]{\@gobble}%
\providecommand \bibinfo  [0]{\@secondoftwo}%
\providecommand \bibfield  [0]{\@secondoftwo}%
\providecommand \translation [1]{[#1]}%
\providecommand \BibitemOpen [0]{}%
\providecommand \bibitemStop [0]{}%
\providecommand \bibitemNoStop [0]{.\EOS\space}%
\providecommand \EOS [0]{\spacefactor3000\relax}%
\providecommand \BibitemShut  [1]{\csname bibitem#1\endcsname}%
\let\auto@bib@innerbib\@empty
\bibitem [{\citenamefont {Gonzalez-Garcia}\ and\ \citenamefont
  {Nir}(2003)}]{GonzalezGarcia:2002dz}%
  \BibitemOpen
  \bibfield  {author} {\bibinfo {author} {\bibfnamefont {M.~C.}\ \bibnamefont
  {Gonzalez-Garcia}}\ and\ \bibinfo {author} {\bibfnamefont {Y.}~\bibnamefont
  {Nir}},\ }\href {\doibase 10.1103/RevModPhys.75.345} {\bibfield  {journal}
  {\bibinfo  {journal} {Rev.Mod.Phys.}\ }\textbf {\bibinfo {volume} {75}},\
  \bibinfo {pages} {345} (\bibinfo {year} {2003})}\BibitemShut {NoStop}%
\bibitem [{\citenamefont {Mention}\ \emph {et~al.}(2011)\citenamefont {Mention}
  \emph {et~al.}}]{Mention:2011rk}%
  \BibitemOpen
  \bibfield  {author} {\bibinfo {author} {\bibfnamefont {G.}~\bibnamefont
  {Mention}} \emph {et~al.},\ }\href {\doibase 10.1103/PhysRevD.83.073006}
  {\bibfield  {journal} {\bibinfo  {journal} {Phys. Rev. D}\ }\textbf {\bibinfo
  {volume} {83}},\ \bibinfo {pages} {073006} (\bibinfo {year}
  {2011})}\BibitemShut {NoStop}%
\bibitem [{\citenamefont {Huber}(2011)}]{Huber:2011wv}%
  \BibitemOpen
  \bibfield  {author} {\bibinfo {author} {\bibfnamefont {P.}~\bibnamefont
  {Huber}},\ }\href {\doibase 10.1103/PhysRevC.84.024617} {\bibfield  {journal}
  {\bibinfo  {journal} {Phys. Rev. C}\ }\textbf {\bibinfo {volume} {84}},\
  \bibinfo {pages} {024617} (\bibinfo {year} {2011})}\BibitemShut {NoStop}%
\bibitem [{\citenamefont {Aguilar-Arevalo}\ \emph {et~al.}(2010)\citenamefont
  {Aguilar-Arevalo} \emph {et~al.}}]{AguilarArevalo:2010wv}%
  \BibitemOpen
  \bibfield  {author} {\bibinfo {author} {\bibfnamefont {A.~A.}\ \bibnamefont
  {Aguilar-Arevalo}} \emph {et~al.} (\bibinfo {collaboration} {MiniBooNE
  Collaboration}),\ }\href {\doibase 10.1103/PhysRevLett.105.181801} {\bibfield
   {journal} {\bibinfo  {journal} {Phys. Rev. Lett.}\ }\textbf {\bibinfo
  {volume} {105}},\ \bibinfo {pages} {181801} (\bibinfo {year}
  {2010})}\BibitemShut {NoStop}%
\bibitem [{\citenamefont {Abazajian}\ \emph {et~al.}(2012)\citenamefont
  {Abazajian} \emph {et~al.}}]{Abazajian:2012ys}%
  \BibitemOpen
  \bibfield  {author} {\bibinfo {author} {\bibfnamefont {K.~N.}\ \bibnamefont
  {Abazajian}} \emph {et~al.},\ }\href@noop {} {\  (\bibinfo {year} {2012})},\
  \Eprint {http://arxiv.org/abs/arXiv:1204.5379} {arXiv:1204.5379} \BibitemShut
  {NoStop}%
\bibitem [{\citenamefont {Ade}\ \emph {et~al.}(2015)\citenamefont {Ade} \emph
  {et~al.}}]{Ade:2015xua}%
  \BibitemOpen
  \bibfield  {author} {\bibinfo {author} {\bibfnamefont {P.~A.~R.}\
  \bibnamefont {Ade}} \emph {et~al.} (\bibinfo {collaboration} {Planck}),\
  }\href@noop {} {\  (\bibinfo {year} {2015})},\ \Eprint
  {http://arxiv.org/abs/1502.01589} {arXiv:1502.01589} \BibitemShut {NoStop}%
\bibitem [{\citenamefont {Hannestad}\ \emph {et~al.}(2014)\citenamefont
  {Hannestad}, \citenamefont {Hansen},\ and\ \citenamefont
  {Tram}}]{Hannestad:2013ana}%
  \BibitemOpen
  \bibfield  {author} {\bibinfo {author} {\bibfnamefont {S.}~\bibnamefont
  {Hannestad}}, \bibinfo {author} {\bibfnamefont {R.~S.}\ \bibnamefont
  {Hansen}}, \ and\ \bibinfo {author} {\bibfnamefont {T.}~\bibnamefont
  {Tram}},\ }\href {\doibase 10.1103/PhysRevLett.112.031802} {\bibfield
  {journal} {\bibinfo  {journal} {Phys.Rev.Lett.}\ }\textbf {\bibinfo {volume}
  {112}},\ \bibinfo {pages} {031802} (\bibinfo {year} {2014})}\BibitemShut
  {NoStop}%
\bibitem [{\citenamefont {Dasgupta}\ and\ \citenamefont
  {Kopp}(2014)}]{Dasgupta:2013zpn}%
  \BibitemOpen
  \bibfield  {author} {\bibinfo {author} {\bibfnamefont {B.}~\bibnamefont
  {Dasgupta}}\ and\ \bibinfo {author} {\bibfnamefont {J.}~\bibnamefont
  {Kopp}},\ }\href {\doibase 10.1103/PhysRevLett.112.031803} {\bibfield
  {journal} {\bibinfo  {journal} {Phys.Rev.Lett.}\ }\textbf {\bibinfo {volume}
  {112}},\ \bibinfo {pages} {031803} (\bibinfo {year} {2014})}\BibitemShut
  {NoStop}%
\bibitem [{\citenamefont {Pospelov}(2011)}]{Pospelov:2011ha}%
  \BibitemOpen
  \bibfield  {author} {\bibinfo {author} {\bibfnamefont {M.}~\bibnamefont
  {Pospelov}},\ }\href {\doibase 10.1103/PhysRevD.84.085008} {\bibfield
  {journal} {\bibinfo  {journal} {Phys.Rev.}\ }\textbf {\bibinfo {volume}
  {D84}},\ \bibinfo {pages} {085008} (\bibinfo {year} {2011})}\BibitemShut
  {NoStop}%
\bibitem [{\citenamefont {Adamson}\ \emph {et~al.}(2013)\citenamefont {Adamson}
  \emph {et~al.}}]{Adamson:2013whj}%
  \BibitemOpen
  \bibfield  {author} {\bibinfo {author} {\bibfnamefont {P.}~\bibnamefont
  {Adamson}} \emph {et~al.} (\bibinfo {collaboration} {MINOS Collaboration}),\
  }\href@noop {} {\bibfield  {journal} {\bibinfo  {journal} {Phys. Rev. Lett.
  110,}\ }\textbf {\bibinfo {volume} {251801}} (\bibinfo {year}
  {2013})}\BibitemShut {NoStop}%
\bibitem [{\citenamefont {Pospelov}(2009)}]{Pospelov:2008zw}%
  \BibitemOpen
  \bibfield  {author} {\bibinfo {author} {\bibfnamefont {M.}~\bibnamefont
  {Pospelov}},\ }\href {\doibase 10.1103/PhysRevD.80.095002} {\bibfield
  {journal} {\bibinfo  {journal} {Phys. Rev.}\ }\textbf {\bibinfo {volume}
  {D80}},\ \bibinfo {pages} {095002} (\bibinfo {year} {2009})}\BibitemShut
  {NoStop}%
\bibitem [{foo({\natexlab{a}})}]{footnote1}%
  \BibitemOpen
  \bibinfo {note} {All the $SU(2)_L$ doublets are assumed to be singlets of
  $U(1)_X$. However, the charged leptons could be connected to the $X$ boson
  thorough $e.g.$ a Higgs loop.}\BibitemShut {Stop}%
\bibitem [{\citenamefont {Dziewonski}\ and\ \citenamefont
  {Anderson}(1981)}]{Dziewonski1981297}%
  \BibitemOpen
  \bibfield  {author} {\bibinfo {author} {\bibfnamefont {A.~M.}\ \bibnamefont
  {Dziewonski}}\ and\ \bibinfo {author} {\bibfnamefont {D.~L.}\ \bibnamefont
  {Anderson}},\ }\href {\doibase
  http://dx.doi.org/10.1016/0031-9201(81)90046-7} {\bibfield  {journal}
  {\bibinfo  {journal} {Physics of the Earth and Planetary Interiors}\ }\textbf
  {\bibinfo {volume} {25}},\ \bibinfo {pages} {297 } (\bibinfo {year}
  {1981})}\BibitemShut {NoStop}%
\bibitem [{\citenamefont {Peres}\ and\ \citenamefont
  {Smirnov}(2001)}]{Peres:2000ic}%
  \BibitemOpen
  \bibfield  {author} {\bibinfo {author} {\bibfnamefont {O.~L.~G.}\
  \bibnamefont {Peres}}\ and\ \bibinfo {author} {\bibfnamefont {A.~Y.}\
  \bibnamefont {Smirnov}},\ }\href {\doibase 10.1016/S0550-3213(01)00012-8}
  {\bibfield  {journal} {\bibinfo  {journal} {Nucl.Phys.}\ }\textbf {\bibinfo
  {volume} {B599}},\ \bibinfo {pages} {3} (\bibinfo {year} {2001})}\BibitemShut
  {NoStop}%
\bibitem [{\citenamefont {Mikheyev}\ and\ \citenamefont {Smirnov}(1985)}]{msw}%
  \BibitemOpen
  \bibfield  {author} {\bibinfo {author} {\bibfnamefont {S.~P.}\ \bibnamefont
  {Mikheyev}}\ and\ \bibinfo {author} {\bibfnamefont {A.~Y.}\ \bibnamefont
  {Smirnov}},\ }\href@noop {} {\bibfield  {journal} {\bibinfo  {journal} {Sov.
  Jour. Nucl. Phys.}\ }\textbf {\bibinfo {volume} {42}},\ \bibinfo {pages}
  {913} (\bibinfo {year} {1985})}\BibitemShut {NoStop}%
\bibitem [{\citenamefont {Gonzalez-Garcia}\ \emph {et~al.}(2012)\citenamefont
  {Gonzalez-Garcia}, \citenamefont {Maltoni}, \citenamefont {Salvado},\ and\
  \citenamefont {Schwetz}}]{GonzalezGarcia:2012sz}%
  \BibitemOpen
  \bibfield  {author} {\bibinfo {author} {\bibfnamefont {M.~C.}\ \bibnamefont
  {Gonzalez-Garcia}}, \bibinfo {author} {\bibfnamefont {M.}~\bibnamefont
  {Maltoni}}, \bibinfo {author} {\bibfnamefont {J.}~\bibnamefont {Salvado}}, \
  and\ \bibinfo {author} {\bibfnamefont {T.}~\bibnamefont {Schwetz}},\
  }\href@noop {} {\bibfield  {journal} {\bibinfo  {journal} {JHEP}\ }\textbf
  {\bibinfo {volume} {12}},\ \bibinfo {pages} {123} (\bibinfo {year}
  {2012})}\BibitemShut {NoStop}%
\bibitem [{foo({\natexlab{b}})}]{footnote2}%
  \BibitemOpen
  \bibinfo {note} {{By marginalizing the $\chi^2$ function of Eq.~(\ref{eq4.7})
  for the 3 neutrino case, we find that our best fit values for
  $\sin^22\theta_{23}$ and $\Delta m^2_{31}$ are fairly close to the values
  reported by the collaboration}}\BibitemShut {NoStop}%
\bibitem [{\citenamefont {Mishra}\ \emph {et~al.}(1991)\citenamefont {Mishra}
  \emph {et~al.}}]{Mishra:1991bv}%
  \BibitemOpen
  \bibfield  {author} {\bibinfo {author} {\bibfnamefont {S.~R.}\ \bibnamefont
  {Mishra}} \emph {et~al.} (\bibinfo {collaboration} {CCFR}),\ }\href {\doibase
  10.1103/PhysRevLett.66.3117} {\bibfield  {journal} {\bibinfo  {journal}
  {Phys.Rev.Lett.}\ }\textbf {\bibinfo {volume} {66}},\ \bibinfo {pages} {3117}
  (\bibinfo {year} {1991})}\BibitemShut {NoStop}%
\bibitem [{\citenamefont {Altmannshofer}\ \emph {et~al.}(2014)\citenamefont
  {Altmannshofer}, \citenamefont {Gori}, \citenamefont {Pospelov},\ and\
  \citenamefont {Yavin}}]{Altmannshofer:2014pba}%
  \BibitemOpen
  \bibfield  {author} {\bibinfo {author} {\bibfnamefont {W.}~\bibnamefont
  {Altmannshofer}}, \bibinfo {author} {\bibfnamefont {S.}~\bibnamefont {Gori}},
  \bibinfo {author} {\bibfnamefont {M.}~\bibnamefont {Pospelov}}, \ and\
  \bibinfo {author} {\bibfnamefont {I.}~\bibnamefont {Yavin}},\ }\href
  {\doibase 10.1103/PhysRevLett.113.091801} {\bibfield  {journal} {\bibinfo
  {journal} {Phys.Rev.Lett.}\ }\textbf {\bibinfo {volume} {113}},\ \bibinfo
  {pages} {091801} (\bibinfo {year} {2014})}\BibitemShut {NoStop}%
\bibitem [{\citenamefont {Davoudiasl}\ \emph {et~al.}(2014)\citenamefont
  {Davoudiasl}, \citenamefont {Lee},\ and\ \citenamefont
  {Marciano}}]{Davoudiasl:2014kua}%
  \BibitemOpen
  \bibfield  {author} {\bibinfo {author} {\bibfnamefont {H.}~\bibnamefont
  {Davoudiasl}}, \bibinfo {author} {\bibfnamefont {H.-S.}\ \bibnamefont {Lee}},
  \ and\ \bibinfo {author} {\bibfnamefont {W.~J.}\ \bibnamefont {Marciano}},\
  }\href {\doibase 10.1103/PhysRevD.89.095006} {\bibfield  {journal} {\bibinfo
  {journal} {Phys. Rev.}\ }\textbf {\bibinfo {volume} {D89}},\ \bibinfo {pages}
  {095006} (\bibinfo {year} {2014})}\BibitemShut {NoStop}%
\bibitem [{\citenamefont {Kopp}\ and\ \citenamefont
  {Welter}(2014)}]{Kopp:2014fha}%
  \BibitemOpen
  \bibfield  {author} {\bibinfo {author} {\bibfnamefont {J.}~\bibnamefont
  {Kopp}}\ and\ \bibinfo {author} {\bibfnamefont {J.}~\bibnamefont {Welter}},\
  }\href {\doibase 10.1007/JHEP12(2014)104} {\bibfield  {journal} {\bibinfo
  {journal} {JHEP}\ }\textbf {\bibinfo {volume} {1412}},\ \bibinfo {pages}
  {104} (\bibinfo {year} {2014})}\BibitemShut {NoStop}%
\bibitem [{\citenamefont {Berryman}\ \emph {et~al.}(2015)\citenamefont
  {Berryman}, \citenamefont {de~Gouvêa}, \citenamefont {Kelly},\ and\
  \citenamefont {Kobach}}]{Berryman:2015nua}%
  \BibitemOpen
  \bibfield  {author} {\bibinfo {author} {\bibfnamefont {J.~M.}\ \bibnamefont
  {Berryman}}, \bibinfo {author} {\bibfnamefont {A.}~\bibnamefont {de~Gouvêa}},
  \bibinfo {author} {\bibfnamefont {K.~J.}\ \bibnamefont {Kelly}}, \ and\
  \bibinfo {author} {\bibfnamefont {A.}~\bibnamefont {Kobach}},\ }\href
  {\doibase 10.1103/PhysRevD.92.073012} {\bibfield  {journal} {\bibinfo
  {journal} {Phys. Rev.}\ }\textbf {\bibinfo {volume} {D92}},\ \bibinfo {pages}
  {073012} (\bibinfo {year} {2015})}\BibitemShut {NoStop}%
\end{thebibliography}%

\end{document}